# Amino Acids Modulate Liquid-Liquid Phase Separation *in vitro* and *in vivo* by Regulating Protein-Protein Interactions

Xufeng Xu[1], Aleksander A. Rebane[2,3], Laura Roset Julia[1], Kathryn A. Rosowski[2], Eric R. Dufresne[2,4], Francesco Stellacci[1,5*]

1: Institute of Materials, Ecole Polytechnique Fédérale de Lausanne (EPFL), Lausanne, 1015, Switzerland

2: Laboratory of Soft and Living Materials, Department of Materials, ETH Zurich, 8093 Zurich, Switzerland

3: Life Molecules and Materials Lab, Programs in Chemistry and Physics, New York University Abu Dhabi, P.O. Box 129188, Abu Dhabi, United Arab Emirates.

4: Department of Materials Science and Engineering, Department of Physics, Cornell University, Ithaca, NY 14853 USA

5: Bioengineering Institute, Ecole Polytechnique Fédérale de Lausanne (EPFL), Lausanne, 1015, Switzerland

E-mail: francesco.stellacci@epfl.ch

**Abstract**: Liquid–liquid phase separation (LLPS) is an intracellular process widely used by cells for many key biological functions. It occurs in complex and crowded environments, where amino acids (AAs) are vital components. We have found that AAs render the net interaction between proteins more repulsive. Here, we find that some AAs efficiently suppress LLPS in test tubes (*in vitro*). We then screen all the proteinogenic AAs and find that three specific AAs, including proline, glutamine, and glycine, significantly suppressed the formation of stress granules (SGs) in U2OS and HeLa cell lines (*in vivo*) irrespective of stress types. We also observe the effect in primary fibroblast cells, a viable cell model for neurodegenerative disorder diseases. Kinetic studies by live-cell microscopy show that the presence of AAs not only slows down the formation but also decreases the saturating concentration and prevents the coalescence of SGs. We finally use sedimentation-diffusion equilibrium analytical ultracentrifuge (SE-AUC) to demonstrate that the suppression effects of AAs on LLPS may be due to their modulation in protein-protein and RNA-RNA interactions. Overall, this study reveals an underappreciated role of cellular AAs, which may find biomedical applications, especially in the treatment of SG-associated diseases.

**Significance statement:** Amino acids (AAs) constitute a major component in intracellular biomolecules. More than 25% of the total volume of a mammalian cell was reported to be made of AAs. Here, we study how AAs affect intracellular liquid–liquid phase separation of proteins, which is relevant to occurrence of neurodegeneration diseases. We find that AAs play an important role in modulating phase separation in test tubes and live cells. Importantly, we find that three specific AAs, including proline, glutamine, and glycine, significantly depress the formation of stress granules in cell lines and primary cells, which can be explained by rendering net homotypic interactions more repulsive. These results deepen our understanding of the under-appreciated roles of cellular AAs in modulating intracellular phase separation.

**Introductions**:

Liquid–liquid phase separation (LLPS) of proteins and nucleic acids(*1–4*) is a versatile intracellular process, which leads to the formation of biomolecular condensates (BMCs). The occurrence of LLPS was shown to be a widely-spread phenomenon in biology(*5*). The detailed LLPS process and the thus-forming BMCs were first revealed by Brangwynne et.al (*6*) by studying P granules in germ cells of *Caenorhabditis elegans*. Subsequently, the fundamental roles of LLPS and BMCs in cellular homeostasis and disease development have been extensively studied in the past decade(*7, 8*). To study the LLPS process and thus forming BMCs(*9*), phase-separation-prone proteins are normally purified and incubated *in vitro*. Jawerth et al.(*10*) used purified proteins from the FUS family and incubated them in a buffer to study the time-dependent rheological properties of BMCs after the LLPS. Wegmann et al.(*11*) used purified intrinsically disordered tau protein *in vitro* to study the LLPS process and the following aggregation process. It is known that *in-vitro* LLPS are quite different from the ones that occur in living cells (*in vivo*)(*4*). There are many known intracellular biomolecules, such as phospholipids(*12*) and ATP(*13*), that have been shown to affect LLPS *in vivo* but are typically not present in *in-vitro* experiments of LLPS.

Indeed, there are many small molecules in the cytosol of a cell. Among them, amino acids (AAs) were reported to constitute a major component in cellular biomolecules. More than 25% of the total volume and 6% of the total mass of a mammalian cell was reported to be made of free AAs(*15*, *16*). The total AA concentration in the cytosol of a mammalian cell was also reported to reach around 50 mM in normal conditions(*17*, *18*) and increase significantly to several hundreds of millimoles in response to stress(*19*, *20*). It is already found that AAs are employed as osmolytes(*21*) to stabilize protein conformation/folding(*22*, *23*) against protein denaturation and aggregation. Yancy et al.(*24*) reported that AAs such as proline and glutamic acid are employed by water-stressed organisms at an intermediately high concentration (hundreds of millimoles) to reduce the aggregation of proteins. Ignatova and Gierasch(*20*) also found that the bacteria *Escherichia coli* regulates the proline concentration to inhibit the initial aggregation process of cellular retinoic acid-binding proteins.

Recently, we have found that proteogenic AAs significantly affect protein-protein interaction by rendering their dispersions more stable, as evidenced by an increase in protein second virial coefficient and a change in protein potential of mean force(*25*). In other words, we have found that AAs, at a concentration range similar to the ones found in the cytosol of mammalian cells (tens to hundreds of millimoles), render proteins *in vitro* more repulsive. Here, we studied the effects of AAs in changing protein-protein interaction and modulating the LLPS in test tubes (*in vitro*) and in cells (*in vivo*). For the *in vitro* studies, *BSA* was chosen as a model folded protein and the low-complexity domain of FUS ($FUS_{267}$) as a model intrinsically disordered protein (IDP). FUS plays a significant role in various cellular processes(*26*) as the mutation of FUS is associated with several diseases, such as amyotrophic lateral sclerosis (ALS) and frontotemporal dementia (FTD)(*27*, *28*). FUS was also reported to be recruited to stress granules (SGs)(*29*) in response to stress. The protein-protein interaction and LLPS of both BSA and $FUS_{267}$ were studied *in vitro* with the presence of proline. We found that the presence of proline stabilized both protein solutions, thus suppressing their phase separation *in vitro*. We further found that proline and various other proteogenic AAs suppressed the formation of SGs in U2OS and HeLa cell lines and the effect is irrespective of stress types (i.e. heat shock or arsenite stress). The effect of AAs on suppressing the formation of SGs was also observed in primary fibroblast cells, a viable cell model for neurodegenerative disorder diseases. The kinetic study by live-cell microscopy experiments showed that the presence of AAs not only slowed down the formation of SGs but also decreased the saturating concentration of SGs and prevented the coalescence of SGs. These modulation effects of AAs on SGs can be explained by weaker net attractive interaction between the components of SGs (including scaffold protein, IDP and RNA) after the addition of AAs. We believe that our study reveals the underappreciated role of cellular AAs on intracellular phase separation.

**Results**:

***In vitro* BSA and FUS$_{267}$ system**. We first studied the effect of proline on bovine serum albumin (BSA) (a folded protein). Sedimentation-diffusion equilibrium analytical ultracentrifuge (SE-AUC) was used to measure the second virial coefficient ($B_{22}$), which quantitatively characterized protein-protein interaction(*32*). The measurement principle is based on the equation of state (**Equation 1**):

$$\frac{\Pi}{kT} = \rho_2 + B_{22}\rho_2^2 + B_{23}\rho_2\rho_3 + \cdots \quad (1)$$

where $\Pi$ is the osmotic pressure, $\rho$ is the number density, $kT$ is the product of the Boltzmann constant by the temperature. $B_{22}$ and $B_{23}$ are both second virial coefficients, where the notation 1 is assigned to the solvent, 2 is assigned to solute 2 (proteins), and 3 to solute 3 (AAs). In this case, $B_{22}$ indicates protein-protein interaction. A positive change ($\Delta B_{22} > 0$) indicates that the protein solution becomes more stable (or more rigorously that the difference between inter-protein repulsion and attraction grows in favor of the former) while $\Delta B_{22} < 0$ indicates that the solution becomes less stable. The procedure to obtain $B_{22}$ is illustrated in **Figure 1A**. As shown in **Figure 1B**, we find that the larger the proline concentration, the larger $\Delta B_{22}$. When no proline is present in the solution, $B_{22}$ for BSA-BSA interaction is $1.0 \times 10^{-24} \pm 2 \times 10^{-26} \ m^3$, corresponding to a BSA diameter of ~ 8 nm if hard-sphere interaction potential is assumed(*33*). This value is in good agreement with the hydrodynamic diameter of BSA (7.8 nm) in water(*34*). With the addition of 1 M proline, the value of $B_{22}$ is increased by up to 40% (from $1.0 \times 10^{-24} \ m^3$ to $1.4 \times 10^{-24} \ m^3$), which indicates a more stable BSA solution. Following a literature procedures(*35*) (experimental details in **Materials and Methods**), polyethylene glycol (PEG 4000 Da) was added into the BSA solution as a crowding agent to induce the phase separation of BSA. The system spontaneously separates into two phases, a BSA-rich droplet phase and a BSA-poor dilute phase. The two phases were separated by centrifugation, **Figure 1C**. UV-vis spectroscopy was used to measure the BSA concentration in the BSA dilute phase. As shown in **Figure 1D**, we found that the presence of proline in the BSA solution increased the BSA concentration in the dilute phase. We also analysed the BSA droplet phase by widefield optical microscopy (experimental details in **Materials and Methods**, 60 mins of incubation time before microscopy experiments to allow for sufficient droplet sedimentation on the sample well bottom) as a function of proline concentration. We found that the size of droplets decreased with higher proline concentrations (**Figure S1A** and **S1B**). The increase in BSA concentration in the dilute phase as well as the decline of the size of the BSA-rich droplets upon the addition of proline indicates a more stable BSA solution.

We also studied the effect of proline on LLPS of the low-complexity domain of fused in sarcoma (FUS$_{267}$, an intrinsically disordered protein(*36*)). As shown in **Figure 1E**, $\Delta B_{22}$ for FUS$_{267}$–FUS$_{267}$ interaction increased with higher proline concentrations (from 0 to 0.2 and 0.5 M), indicating a more stable FUS$_{267}$ solution after the addition of proline. The LLPS of FUS$_{267}$ was induced according to the literature procedure(*36*) (experimental details in **Materials and Methods**). As shown in **Figure 1F**, the FUS$_{267}$ concentration in the dilute phase increased with higher proline concentrations, as measured via UV-VIS spectroscopy. By using widefield optical microscopy, we also observed that the smaller the size of the FUS$_{267}$ droplets, the higher the proline concentration (**Figure 1G**, the related widefield optical microscopy images shown in **Figure S1C**).

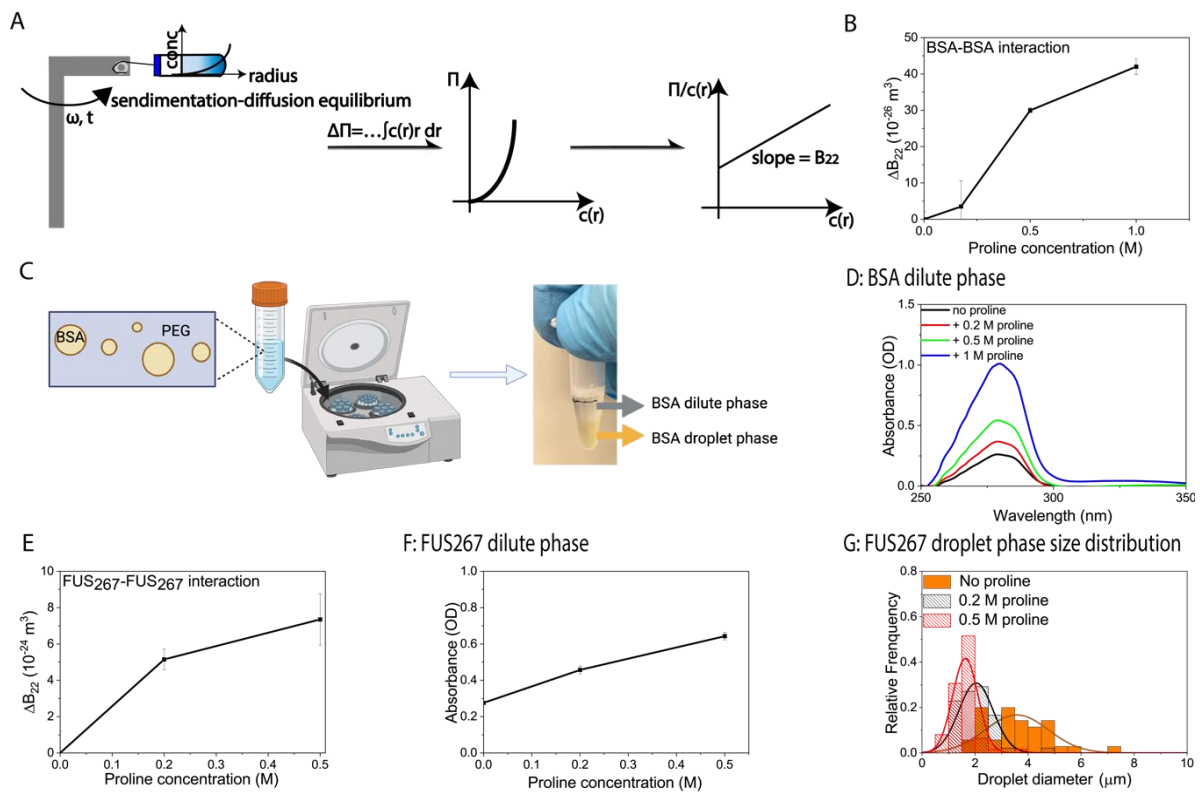

**Figure 1**: **The effect of proline in BSA-BSA interaction and phase separation, and FUS$_{267}$- FUS$_{267}$ interaction and phase separation**. A. Schematic representation of the procedures for the calculation of $B_{22}$ by SE-AUC; B. The change of $B_{22}$ ($\Delta B_{22}$) for BSA-BSA interaction with different proline concentrations in BSA solution(*32*); C. Schematic representation of the procedures for the separation of the BSA dilute and droplet phases by centrifugation after the spontaneous phase separation of BSA with 4000 Da polyethylene glycol (PEG) as a crowding agent; D. UV-vis spectroscopy of the BSA dilute phase with different proline concentrations; E. Change of $B_{22}$ ($\Delta B_{22}$) for FUS$_{267}$–FUS$_{267}$ interaction

with different proline concentrations (0.2 and 0.5 M) in FUS$_{267}$ solution; F. Absorbance at 275 nm of the FUS$_{267}$ dilute phase as a function of proline concentrations (0, 0.2 and 0.5 M); G. FUS$_{267}$ droplet size distribution as a function of proline concentrations (0, 0.2 and 0.5 M).

***In vivo* U2OS and HeLa cell lines**. After the *in vitro* experiments with BSA and FUS$_{267}$, we studied the effect of proline on the intracellular LLPS *in vivo* by employing stress granules (SGs)(*31, 37*), a well-known example of biomolecular condensates (BMCs) that form via the phase separation of a diverse set of proteins and RNAs. In this study, the SGs in U2OS cells were induced in response to oxidative stress by the addition of sodium arsenite (NaAs). It is noteworthy that SGs contain a variety of client proteins, including FUS(*38*). In the positive control row of **Figure 2A**, we show fluorescence microscopy images of fixed U2OS cells that developed SGs after the oxidative stress for 45 mins (indicated by the arrowheads in the alexa fluor channel). We pre-treated the cells with 200 mM proline for 2 h before the stress. After the pre-treatment, the intracellular proline concentration increased to ~ 130 mM, as estimated by the LC-MS, **Figure S2** (the details of intracellular amino acid extraction protocol and LC-MS quantification analysis are described in **SI1**). We believe that the significant proline concentration increase in the cytosol is due to facilitated diffusion(*39*) of AAs, a process of spontaneous passive transport of AAs across the cell membrane via specific transmembrane integral protein transporters(*40*). Cell viability after the pre-treatment with proline was also measured by the cell proliferation assay (MTT assay), as described in **SI2**. It was found that the viability of the cells did not change after the uptake of proline (**Figure S3**). NaAs was then added to the cells and incubated for ~ 45 mins at room temperature to induce the formation of SGs (the details of cell culture, cell stress, and fixing procedures described in **Materials and Methods**). As shown in **Figure 2A**, we observed that the number of SGs in U2OS cells decreased with the pre-treatment of 200 mM proline compared to the positive control condition (the treatment of only 200 mM proline without the later addition of NaAs did not change the distribution of G3BP in the cytosol, **Figure S4**). We further quantified the number change of SGs per cell after the proline pre-treatment (detailed analysis method and script described in **Materials and Methods**). As shown in **Figure 2B**, we found that the pre-treatment of proline at concentrations larger than 200 mM leads to a statistically significant decrease in the average number of SGs per cell (P < 0.0001). We then screened all the other neutral proteinogenic AAs (with a solubility higher than 100 mM) for their effects on SGs and we found that the pre-treatment with glutamine (200 mM) and glycine (400 mM) also decreased the SG number significantly (P < 0.0001), **Figure 2C**. The pre-treatment with serine, valine, and alanine at 400 mM did not lead to any significant change in the average number of SGs per cell but led to a significant decrease in the average size of SGs, **Figure S5**. Charged proteinogenic AAs were also screened. As shown in **Figure S6**, negatively charged asparagine (20 mM) decreased both the number of SGs (P < 0.01) and the size of SGs (P < 0.1) while positively charged histidine (100 mM) decreased the size of SGs (P < 0.01). We also observed that the presence of AAs prevents the formation of SGs in response to heat shock stress, **Figure 2A**. As shown in **Figure 2D**, proline, glutamine, and glycine all significantly decreased the number of SGs per cell after the heat shock treatment at 43 ºC, 5% $CO_2$ for 1 h. The effect of proline on the average number of SGs per cell was also tested in HeLa cells and the same phenomenon was observed (**Figure 2E**). Proline, glutamine, and glycine all significantly decreased the average number of SGs per cell in HeLa cells (P < 0.0001), Figures **2F** and **G.** Apart from proline (**Figure 2B**), the effect of glutamine and glycine on the average number of SGs per cell was also found to be concentration dependent (**Figure S7**).

***In vivo* primary fibroblast cells.** To address the relevance of our findings of the effects of AAs on SGs to disease treatment, we performed experiments using primary human dermal fibroblast cells, a viable cell model for studying neurodegenerative disorders associated with SGs(*41*, *42*). Many SGs formed in the cytosol of the primary fibroblast cells (indicated by arrowheads) after the incubation with NaAs (1 mM) for 2 h at RT (positive control, **Figure 3B**), compared to no SG formation in the negative control (**Figure 3A**). With the pre-treatment of proline (200 mM), glutamine (200 mM), and glycine (400 mM) before the NaAs stress, the formation of SGs was significantly suppressed (**Figure 3C**, **D,** and **E**). The average number of SGs per cell in fibroblast cells was significantly decreased after the pre-treatment with these AAs (P < 0.0001), **Figure 3F**.

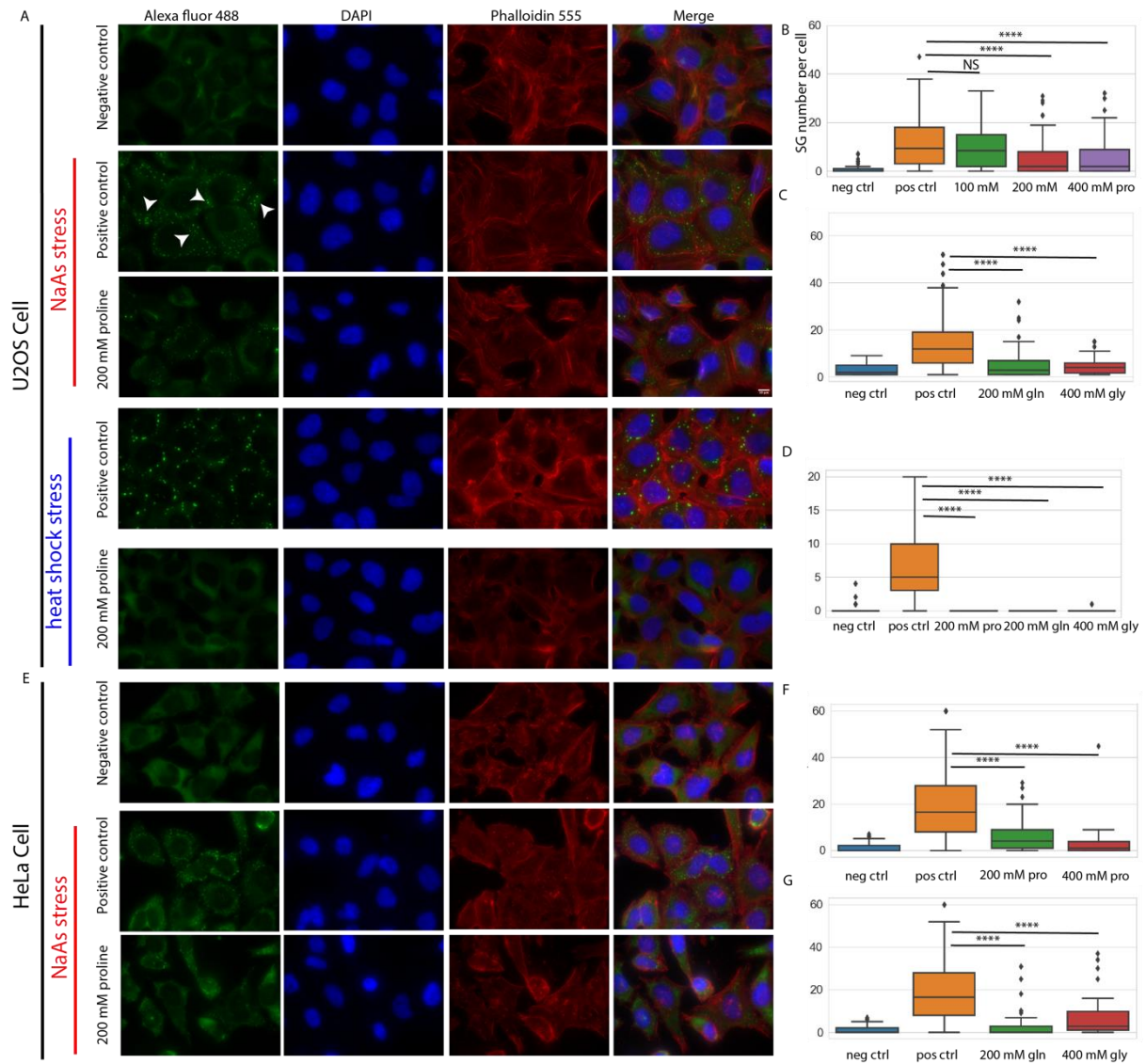

**Figure 2: Immunofluorescence microscopy showing the effect of proline and other neutral amino acids (AAs) on stress granules (SGs) in U2OS and HeLa cells**. A) U2OS cells exhibited a significantly reduced number of SGs following the pre-treatment with 200 mM proline for 2 hours compared to positive control where no exogenous AAs were added in response to NaAs and heat shock stress; B) The number of SGs per cell after the pre-treatment with 100, 200, and 400 mM proline in response to NaAs stress; C) The number of SGs per cell after the pre-treatment with other neutral proteinogenic AAs including glutamine (200 mM), glycine (400 mM) in response to NaAs stress; D) The number of SGs per cell after the pre-treatment with 200 mM proline, 200 mM glutamine and 400 mM glycine in response to heat shock stress; E) HeLa cells exhibited a significantly reduced number of SGs following the pre-treatment with 200 mM proline for 2 hours compared to positive control where no exogenous AAs were added in response to NaAs stress. F) The number of SGs per cell after the pre-treatment with 200 and 400 mM proline in response to NaAs stress; G) The number of SGs per cell after the pre-treatment with glutamine (200 mM), and glycine (400 mM) in response to NaAs stress. (****: P value < 0.0001; NS: no significance, P value > 0.05). Arrowheads: SGs. Scale bar: 10 μm

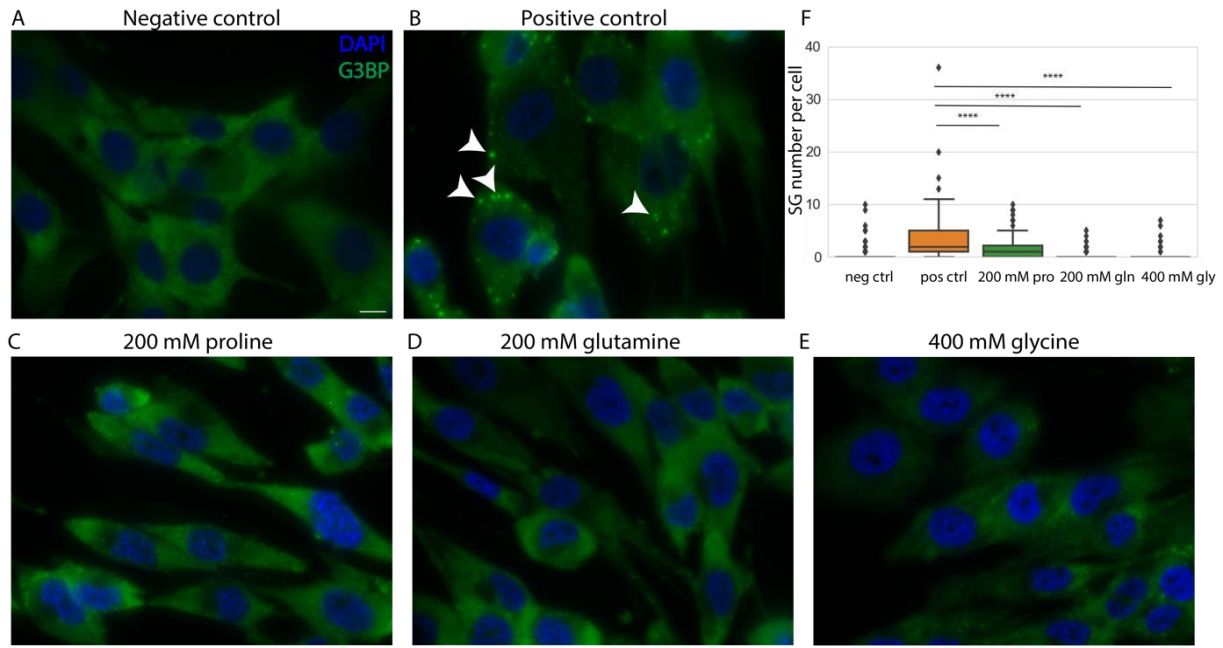

**Figure 3: Immunofluorescence microscopy showing the effect of proline and other neutral amino acids (AAs) on stress granules (SGs) in primary fibroblast cells**. A and B) fibroblast cells exhibited a significant number of SGs following the incubation with NaAs (1 mM) for 2 h at RT; C, D and E) primary fibroblast cells exhibited a significantly reduced number of SGs following the pre-treatment with 200 mM proline (C), 200 mM glutamine (D), 400 mM glycine (E) for 2 h; F) The number of SGs per cell after the pre-treatment with proline (200 mM), glutamine (200 mM), and glycine (400 mM) (****: P value < 0.0001). Scale bar: 10 μm. Arrowheads: SGs.

**Formation kinetics of SGs**. To reveal the effect of AAs on the formation process kinetics of SGs, we performed live-cell microscopy of U2OS cells constitutively expressing GFP-tagged G3BP1, a marker protein of SGs(*43, 44*) with the presence of proline and glutamine. As shown in the control set of **Figure 4A** (qualitatively) and **4B** (quantitatively), SGs started to form at 15 mins and saturate at 25 mins, after which they started to coalesce, leading to a decrease in the number of SGs (indicated as lag, growth, and coalescence phases in **Figure 4B**). The coalescence events are indicated inside the red squares in **Figure 4A**. In contrast, the proline pre-treatment for 2 h led to a decrease in the SG formation rate. In addition, the saturating number of SGs per cell decreased by more than half (from 20 to 8 SGs per cell) and the coalescence event became trivial as no significant decrease in the number of SGs was observed. When the proline was added at the same time as NaAs, we also observed a lower rate of SG formation, decreased saturating number of SGs, and diminished coalescence of SGs. However, the effect was less pronounced when compared to the 2 h pre-treatment with proline. This may be because the diffusion process of proline into the cells took time to reach equilibrium. When we post-treated the cells with proline 15 mins after the addition of NaAs, we observed an immediate suppression of the SG formation rate. The SG number saturation was reached at 40 mins, which is 15 mins later than the control set. 200 mM glutamine also showed similar effects, as shown in **Figure 4C** (related live cell images in **Figure S8**). It is noteworthy that the pre-treatment with 200 mM glutamine also delayed the formation of SGs. This delay effect however was not found in the co-treatment and post-treatment experiments.

**Interactions for the formation of SGs**. In light of the finding that proline stabilizes the protein-protein interactions in solution for the low-complexity domain of FUS which is commonly recruited in SGs(*45*) (**Figure 1E**), we hypothesize that AAs suppress the formation of SGs because the presence of AAs renders the net interaction for SG components more repulsive. To prove that, we studied the effects of proline on G3BP-G3BP interaction (key scaffold protein for the formation of SGs(*46*)) as well as ssDNA-ssDNA interaction (nucleic acid to model a large variety of RNA recruited in SGs). As shown in **Figure 5A**, $B_{22}$ for G3BP-G3BP interaction in solution (25 mM Tris-HCl buffer at pH 7.6) increased with higher proline concentrations (from 0 to 0.2 and 0.4 M), indicating a less net G3BP attractive interaction after the addition of proline. As shown in **Figure 5B**, $B_{22}$ for ssDNA-ssDNA interaction in solution (1x TE buffer at pH 7.0) also increased with higher proline concentrations (from 0 to 0.2 and 0.4 M), indicating a more net ssDNA repulsive interaction after the addition of proline.

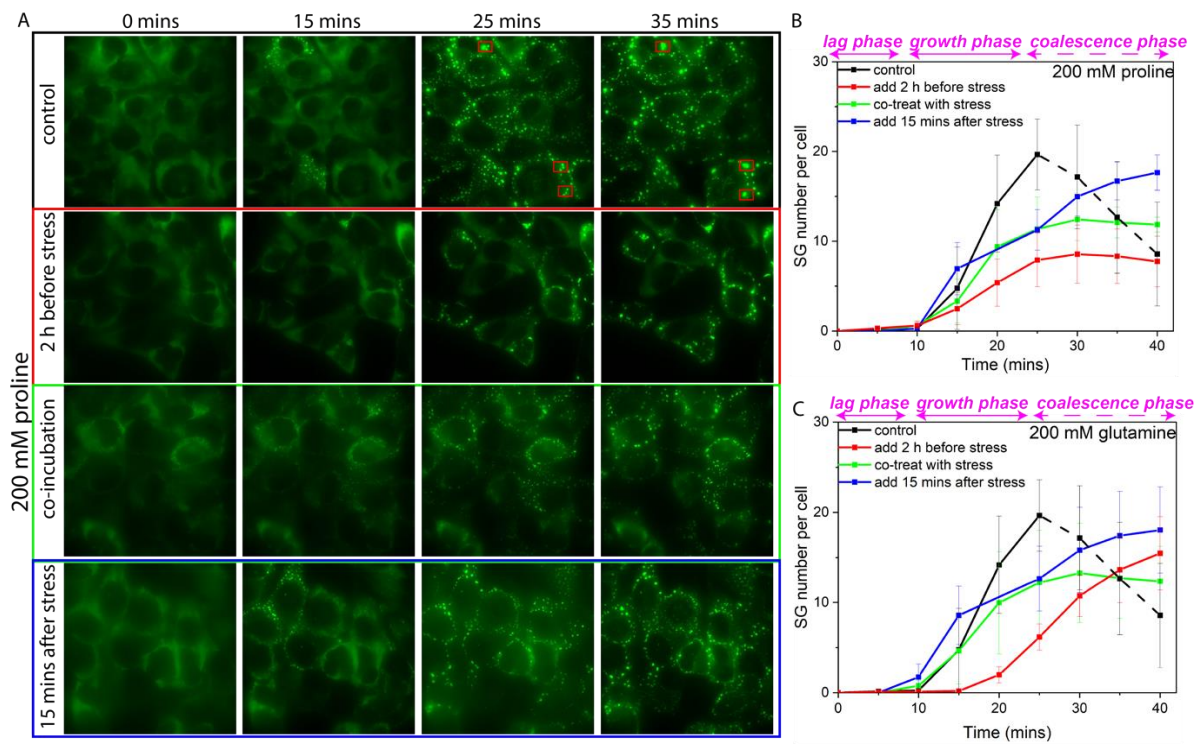

**Figure 4**: **Live cell microscopy of GFP-tagged G3BP1 showing the effect of proline and glutamine (200 mM) on the formation process of stress granules (SGs) in U2OS cells.** A) The pre-treatment for 2 hours, the co-treatment, and the post-treatment (15 mins after the addition of NaAs) with 200 mM proline all suppressed the formation of SGs compared to positive control (no exogenous AAs were added) by live cell microscopy of U2OS cells with GFP tagged G3BP1 (the coalescence events indicated in the red squares); B) time series of the number of SGs per cell after the addition of NaAs including positive control (no proline treatment) and 200 mM proline pre/co-/post-treatment (indicated as lag, growth and coalescence phases). C) time series of the number of SGs per cell after the addition of NaAs including positive control (no glutamine treatment) and 200 mM glutamine pre/co-/post-treatment (related live cell images in **Figure S8**).

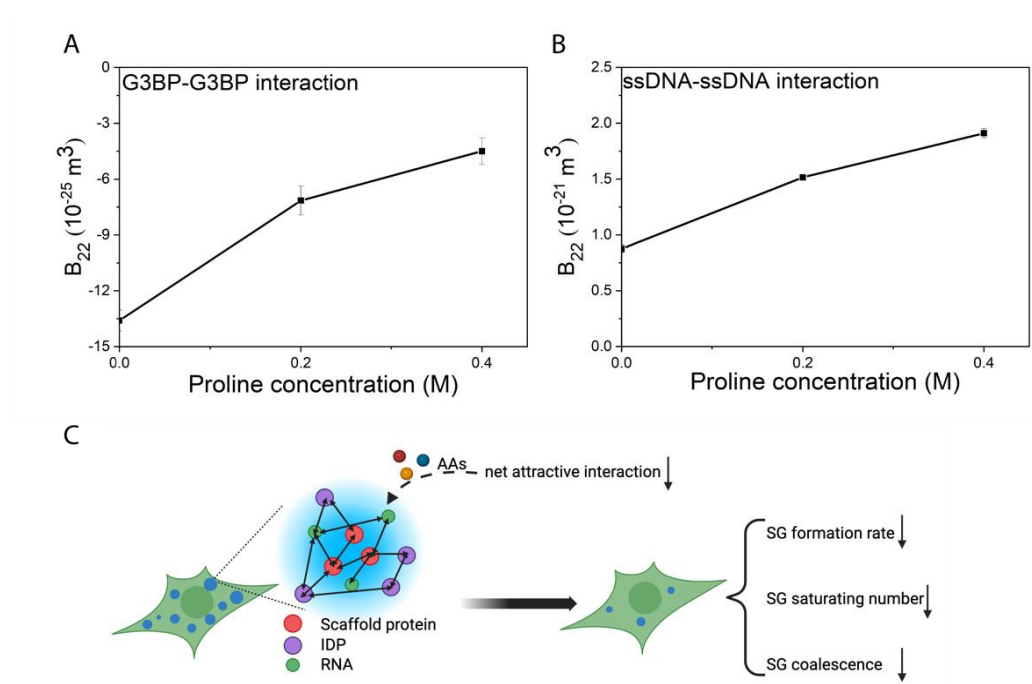

**Figure 5**: **The effect of proline in G3BP-G3BP interaction (A) and ssDNA-ssDNA interaction (B) in solution and the proposed mechanism for the modulation of AAs on the formation of SGs (C).** The SGs are formed in the cytosol as a result of complex interactions among an assembly of proteins (including scaffold proteins which are crucial for the formation of SGs and a large variety of IDP) as well as many RNA in response to stress. The presence of AAs is proposed to render net attractive forces between these SG components weaker and thus modulate the formation of SGs. The modulation effect is composed of three aspects: i) lower SG formation rate; ii) smaller SG saturating number; and iii) less SG coalescence.

**Discussion and Conclusions**:

We find that proline stabilizes the interaction for both BSA (a folded protein) and $FUS_{267}$ (an intrinsically disordered protein) in solution. This is established by our findings that $B_{22}$ in both cases becomes larger with a higher proline concentration. This is consistent with our recent finding that AAs have a general role in stabilizing protein dispersions(*25*). The increase in $B_{22}$ directly implies an increase in the osmotic pressure in the solution, which in turn implies a decrease in the protein chemical potential. An obvious consequence of this latter fact is that proteins' tendency to phase separate changes. Indeed, we have shown here that the addition of proline diminished the LLPS of both BSA and $FUS_{267}$ *in vitro*. The data presented here show a clear effect at a concentration of 200 mM, which is approximately 4 times higher than the typical intracellular amino acid concentration in normal conditions (~50 mM) but at the same concentration range (several hundreds of mM) that is found when osmotic stress is applied to cells(*24*).

Because of these findings, we ventured into investigating the effect of intracellular amino acids on the formation of SGs in living cells (*in vivo*). We found that a variety of neutral AAs at concentrations of hundreds of millimolar can either reduce the average number (proline, glutamine, and glycine) or average size (serine, valine, and alanine) of SGs in two cell lines (i.e. U2OS and HeLa cells). We also found the modulation effect of AAs on SGs of primary fibroblast cells, which are associated with several neurodegenerative diseases, such as Alzheimer's disease (AD), Parkinson's disease (PD), Huntington's disease (HD), and amyotrophic lateral sclerosis (ALS). Furthermore, live cell microscopy study of proline and glutamine revealed the multiple effects of these AAs on the SG formation process: i) they reduce the SG formation rate; ii) they decrease the SG saturating number; and iii) they prevent SG coalescence. These modulation effects of AAs on SGs can be explained by the lower net attractive force between SG components upon the addition of AAs (**Figure 5C**). We corroborated this hypothesis by studying the effects of proline on G3BP-G3BP interaction (key scaffold protein for the formation of SGs) as well as ssDNA-ssDNA interaction (nucleic acid to model a large variety of RNA recruited in SGs). In our recent study of protein interaction stabilization of AAs(*25*), we have established that short peptides are at least as effective as AAs in this process. Given that the cell is also rich in peptides, we will further study whether they can also modulate the LLPS *in vivo* in the next step of our research, which could be connected with the metabolic cycle of cells. We believe that this improved understanding of AAs' effects in tuning intracellular protein phase separation may provide new knowledge of intracellular complex biophysics(*4*, *7*, *48*) and provide a new strategy for the treatment of neurodegenerative disorders associated with stress granules.

**Materials and Methods**:

**Materials**

Bovine serum albumin (BSA) (A7638), polyethylene glycol (PEG) 4 kDa (A16151), and all the amino acids that were used in this study were purchased from Sigma-Aldrich. Recombinant Human G3BP protein was purchased from Abcam (ab103304). FUS low complexity domain (LCD) (residues 1–267) was kindly provided by Dufresne Lab(*36*). U2OS human osteosarcoma cells (wild type) were kindly provided by Dufresne Lab(*43*). U2OS cells stably expressing GFP-tagged G3BP1 were kindly provided by Pelkmans Lab(*44*). HeLa cells (CCL-2™) and Primary Dermal Fibroblast Normal; Human, Neonatal (PCS-201-010™) were purchased from ATCC. ssDNA (p4844) was purchased from Tilibit Nanosystems GmbH.

**SE-AUC experiments**

In a typical experiment, a protein solution in phosphate buffer (pH 7, 50 mM) at a typical concentration of 10 mg/ml was mixed with proline at a series of concentrations (0, 0.2, 0.5, and 1 M). The final solutions were added into the AUC cells (3 mm pathlength). The sedimentation-diffusion equilibria at a proper angular velocity, 20 °C were reached for these samples typically after 24 h. The protein concentration gradient was obtained and converted into the equation of state (EOS) curve (**Equation 1**) using the well-known relation(*49, 50*) (**Equation 2**), where $c$ the mass concentration of the solute species, $\omega$ the angular velocity, $d$ the density and $r$ the radial position, respectively.

$$\Delta \Pi = \omega^2 \left(\frac{\partial d}{\partial c}\right)_\mu \int_{r_m}^{r_1} c(r)\, r\, dr \quad (2)$$

The final step is to calculate $\frac{\Pi}{\rho_2}$, as shown in **Equation 3** (rearranged from **Equation 1**). The slope of the curve ($\frac{\Pi}{\rho_2}$ vs $\rho_2$) equals the value of $B_{22}$.

$$\frac{\Pi}{\rho_2} = 1 + B_{22}\rho_2 + \cdots \quad (3)$$

The change of $B_{22}$ ($\Delta B_{22}$) with the proline concentration series is thus calculated.

**Droplet formation *in vitro***

*BSA droplet*: The method for BSA droplet formation is adjusted from the work by Testa et. al(*35*). Briefly, 100 µL MilliQ water (with different concentrations of proline), 125 µL of 4× potassium phosphate buffer, and 75 µL of 20% (w/v) BSA were mixed in a 1.5 mL Eppendorf tube. Then 200 µL of 60% (w/v) PEG 4K was added to make BSA droplet suspension. The droplet suspension was further stabilized for 60 mins (to allow for sufficient droplet sedimentation to the sample well bottom) and the imaging was done quickly on the bottom of the sample well by Olympus Cell xCellence microscope, using ×60 oil objective with a numerical aperture (NA) of 1.35 and a Hamamatsu ORCA 03g camera. The BSA dilute phase was further separated for UV-vis spectroscopy after centrifugation at 10000 g for 10 minutes.

*FUS droplet*: The method for FUS droplet formation is adjusted from the work by Ijavi et. al(*36*). Briefly, the 2 mM FUS stock (more specifically, FUS low complexity domain (LCD) (residues 1–267) with 6M urea in the buffer (50 mM HEPES and 150 mM NaCl) was diluted 200 times with the buffer (with different concentrations of proline) to induce the phase separation. The droplet suspension was further stabilized for 60 mins and the imaging was done quickly on the bottom of the sample well by Olympus Cell xCellence microscope, using ×60 oil objective with a numerical aperture (NA) of 1.35 and a Hamamatsu ORCA 03g camera. The FUS concentration in the dilute phase was measured by measuring the absorbance at 275 nm in an analytical ultracentrifuge (at 9000 rpm). The droplet size distribution was measured by a simple MATLAB script. In the script, the diameter of a droplet is measured by clicking 2 points on the droplet sphere.

**Cell culture**

U2OS human osteosarcoma cells, HeLa cells (CCL-2 ™), and U2OS cells stably expressing GFP-tagged G3BP1 were grown in Dulbecco's modified eagle medium (high glucose, GlutaMAX™ supplement), supplemented with 5% or 10% fetal bovine serum at 37 °C in 5% $CO_2$. Primary Dermal Fibroblast cells were grown in Fibroblast Growth Medium (C-23110 from PromoCell) at 37 °C in 5% $CO_2$.

**Immunofluorescence microscopy**

After the treatment with 1 mM sodium arsenite (Sigma-Aldrich) for ~ 45 mins at RT (arsenite stress) or at 43 °C, 5% $CO_2$ for 1 h (heat shock stress) to induce stress granule formation, wild-type U2OS cells cultured in μ-Slide 18 Well Glass Bottom (ibid, 81817) were fixed with 4% formaldehyde for 15 min, permeabilized with 0.25% Triton X-100 in phosphate-buffered saline (PBS) for 15 mins and blocked with 20 mg/ml bovine serum albumin (Sigma-Aldrich) in 0.25% Triton X-100. The fixed cells were firstly incubated overnight at 4 °C with primary antibodies of mouse anti-G3BP (1:300, Abcam) and then incubated for 1 h at RT with the secondary antibodies of Alexa fluor 488 anti-mouse IgG (1:300, Abcam). Then, DNA was stained by the incubation in 4′,6-diamidino-2-phenylindole (DAPI) (1:1000, Sigma) for 15 min at RT and F-actin was stained by the incubation in Alexa Fluor™ 555 Phalloidin (1:500, Fischer Scientific). The fixed cells were imaged by Olympus Cell xCellence microscope, using ×60 oil objective with a numerical aperture (NA) of 1.35 and a Hamamatsu ORCA 03g camera.

**Live cell imaging**

Single U2OS cells stably expressing GFP-tagged G3BP1 were cultured in μ-Slide 18 Well Glass Bottom (ibid, 81817) and allowed for growth overnight. The cells were then switched into live-cell imaging media (FluoroBrite™ DMEM with 10% FBS, 10 mM DL-Lactic acid sodium, and ProLong™ Live Antifade Reagent (1:100)) and stabilized at 37 °C in 5% $CO_2$ for 2 h. The cells were imaged at 37 °C and 5% $CO_2$ after 0.5 mM sodium arsenite was added. Imaging was done on an Olympus Cell xCellence microscope with a cage incubator (37 °C and 5% $CO_2$), using ×60 oil objective with a numerical aperture (NA) of 1.35 and an Andor Ixon 3 camera.

**Stress granule and cell detection**

Image analysis of the fixed and live images was performed using Fiji(*51*) and custom scripts. Briefly, cells were segmented using cellpose(*52, 53*) and the stress granules were detected using a fixed threshold or local maxima respectively for the fixed and live images. The optimal values for the fixed threshold or local maxima were obtained when 1) a minimal change was detected in the granule number on varying the threshold value and 2) most of the granules were captured without a minimal number of artificial granules. Scripts and datasets are available at DOI: 10.5281/zenodo.8081112.

**Supporting Information:**

Supporting Information is available. All the metadata are available at DOI: 10.5281/zenodo.10825309.

**Acknowledgements:**

X.X. and F.S. acknowledge the support of the European Union's Horizon 2020 Research and Innovation program under grant agreement no.101017821 (LIGHT-CAP). A.A.R. acknowledges the support of the Ambizione Grant 202214 from the Swiss National Science Foundation. We would like to thank the EPFL BioImaging & Optics Core Facility and in particular José Artacho for his assistance in imaging and Dr. Romain Guiet for his assistance in image processing.

**Author contributions:**

Concept and design: X.X. and F.S.; SE-AUC experiments: X.X.; *in vitro* droplet experiments: X.X. and A.A.R.; *in vivo* cell experiments: X.X. and K.A.R.; lysate extraction and LC-MS: X.X. and L.R.J; microscopy experiments: X.X.; Data analysis and discussion: X.X., F.S. and E.R.D.; Paper writing: X.X. and F.S. with the input from A.A.R., E.R.D. and L.R.J.

**Competing interests:**

The authors declare no competing interests.

**References:**

1. N. A. Yewdall, A. A. M. André, T. Lu, E. Spruijt, Coacervates as models of membraneless organelles. *Current Opinion in Colloid & Interface Science* **52**, 101416 (2021).

2. Z. Gao, W. Zhang, R. Chang, S. Zhang, G. Yang, G. Zhao, Liquid-Liquid Phase Separation: Unraveling the Enigma of Biomolecular Condensates in Microbial Cells. *Frontiers in Microbiology* **12** (2021).

3. A. A. M. André, E. Spruijt, Liquid–Liquid Phase Separation in Crowded Environments. *IJMS* **21**, 5908 (2020).

4. S. Alberti, A. Gladfelter, T. Mittag, Considerations and Challenges in Studying Liquid-Liquid Phase Separation and Biomolecular Condensates. *Cell* **176**, 419–434 (2019).

5. E. Astoricchio, C. Alfano, L. Rajendran, P. A. Temussi, A. Pastore, The Wide World of Coacervates: From the Sea to Neurodegeneration. *Trends in Biochemical Sciences* **45**, 706–717 (2020).

6. C. P. Brangwynne, C. R. Eckmann, D. S. Courson, A. Rybarska, C. Hoege, J. Gharakhani, F. Jülicher, A. A. Hyman, Germline P Granules Are Liquid Droplets That Localize by Controlled Dissolution/Condensation. *Science* **324**, 1729–1732 (2009).

7. S. Alberti, A. A. Hyman, Biomolecular condensates at the nexus of cellular stress, protein aggregation disease and ageing. *Nat Rev Mol Cell Biol* **22**, 196–213 (2021).

8. S. F. Banani, H. O. Lee, A. A. Hyman, M. K. Rosen, Biomolecular condensates: organizers of cellular biochemistry. *Nat Rev Mol Cell Biol* **18**, 285–298 (2017).

9. M. Ijavi, "Physical Characterization of Phase Separated Biomolecular Condensates," thesis, ETH Zurich (2021).

10. L. Jawerth, E. Fischer-Friedrich, S. Saha, J. Wang, T. Franzmann, X. Zhang, J. Sachweh, M. Ruer, M. Ijavi, S. Saha, J. Mahamid, A. A. Hyman, F. Jülicher, Protein condensates as aging Maxwell fluids. *Science* **370**, 1317–1323 (2020).

11. S. Wegmann, B. Eftekharzadeh, K. Tepper, K. M. Zoltowska, R. E. Bennett, S. Dujardin, P. R. Laskowski, D. MacKenzie, T. Kamath, C. Commins, C. Vanderburg, A. D. Roe, Z. Fan, A. M. Molliex, A. Hernandez-Vega, D. Muller, A. A. Hyman, E. Mandelkow, J. P. Taylor, B. T. Hyman, Tau protein liquid–liquid phase separation can initiate tau aggregation. *The EMBO Journal* **37**, e98049 (2018).

12. J. G. Dumelie, Q. Chen, D. Miller, N. Attarwala, S. S. Gross, S. R. Jaffrey, Biomolecular condensates create phospholipid-enriched microenvironments. *Nat Chem Biol*, 1–12 (2023).

13. A. Patel, L. Malinovska, S. Saha, J. Wang, S. Alberti, Y. Krishnan, A. A. Hyman, ATP as a biological hydrotrope. *Science* **356**, 753–756 (2017).

14. I. Piazza, K. Kochanowski, V. Cappelletti, T. Fuhrer, E. Noor, U. Sauer, P. Picotti, A Map of Protein-Metabolite Interactions Reveals Principles of Chemical Communication. *Cell* **172**, 358-372.e23 (2018).

15. R. Rollin, J.-F. Joanny, P. Sens, Physical basis of the cell size scaling laws. *eLife* **12**, e82490 (2023).

16. B. Alberts, *Molecular Biology of the Cell* (Garland Science, 2017).

17. G. Gauthier-Coles, J. Vennitti, Z. Zhang, W. C. Comb, S. Xing, K. Javed, A. Bröer, S. Bröer, Quantitative modelling of amino acid transport and homeostasis in mammalian cells. *Nat Commun* **12**, 5282 (2021).

18. J. Bergström, P. Fürst, L. O. Norée, E. Vinnars, Intracellular free amino acid concentration in human muscle tissue. *Journal of Applied Physiology* **36**, 693–697 (1974).

19. M. T. Fisher, Proline to the rescue. *Proceedings of the National Academy of Sciences* **103**, 13265–13266 (2006).

20. Z. Ignatova, L. M. Gierasch, Inhibition of protein aggregation in vitro and in vivo by a natural osmoprotectant. *Proceedings of the National Academy of Sciences* **103**, 13357–13361 (2006).

21. S. H. Khan, N. Ahmad, F. Ahmad, R. Kumar, Naturally occurring organic osmolytes: From cell physiology to disease prevention. *IUBMB Life* **62**, 891–895 (2010).

22. A. E. Rydeen, E. M. Brustad, G. J. Pielak, Osmolytes and Protein–Protein Interactions. *J. Am. Chem. Soc.* **140**, 7441–7444 (2018).

23. T. O. Street, D. W. Bolen, G. D. Rose, A molecular mechanism for osmolyte-induced protein stability. *Proceedings of the National Academy of Sciences* **103**, 13997–14002 (2006).

24. P. H. Yancey, M. E. Clark, S. C. Hand, R. D. Bowlus, G. N. Somero, Living with Water Stress: Evolution of Osmolyte Systems. *Science* **217**, 1214–1222 (1982).

25. Xufeng Xu, Ting Mao, Pamina M. Winkler, Francesco Stellacci., Amino Acids Effect on Colloidal Properties [Unpublished Manuscript].

26. A. Yamaguchi, K. Takanashi, FUS interacts with nuclear matrix-associated protein SAFB1 as well as Matrin3 to regulate splicing and ligand-mediated transcription. *Sci Rep* **6**, 35195 (2016).

27. O. Broustal, A. Camuzat, L. Guillot-Noël, N. Guy, S. Millecamps, D. Deffond, L. Lacomblez, V. Golfier, D. Hannequin, F. Salachas, W. Camu, M. Didic, B. Dubois, V. Meininger, I. L. Ber, A. Brice, the French clinical and genetic research network on FTD/FTD-MND, FUS Mutations in Frontotemporal Lobar Degeneration with Amyotrophic Lateral Sclerosis. *Journal of Alzheimer's Disease* **22**, 765–769 (2010).

28. T. J. Kwiatkowski, D. A. Bosco, A. L. LeClerc, E. Tamrazian, C. R. Vanderburg, C. Russ, A. Davis, J. Gilchrist, E. J. Kasarskis, T. Munsat, P. Valdmanis, G. A. Rouleau, B. A. Hosler, P. Cortelli, P. J. de Jong, Y. Yoshinaga, J. L. Haines, M. A. Pericak-Vance, J. Yan, N. Ticozzi, T. Siddique, D. McKenna-Yasek, P. C. Sapp, H. R. Horvitz, J. E. Landers, R. H. Brown, Mutations in the FUS/TLS Gene on Chromosome 16 Cause Familial Amyotrophic Lateral Sclerosis. *Science* **323**, 1205–1208 (2009).

29. A. Aulas, C. Vande Velde, Alterations in stress granule dynamics driven by TDP-43 and FUS: a link to pathological inclusions in ALS? *Frontiers in Cellular Neuroscience* **9** (2015).

30. S. P. Plassmeyer, A. S. Holehouse, Stress granules offer first aid for leaky organelles. *Nature*, doi: 10.1038/d41586-023-03417-4 (2023).

31. H. Glauninger, C. J. Wong Hickernell, J. A. M. Bard, D. A. Drummond, Stressful steps: Progress and challenges in understanding stress-induced mRNA condensation and accumulation in stress granules. *Molecular Cell* **82**, 2544–2556 (2022).

32. X. Xu, Q. Ong, T. Mao, P. J. Silva, S. Shimizu, L. Rebecchi, I. Kriegel, F. Stellacci, Experimental Method to Distinguish between a Solution and a Suspension. *Advanced Materials Interfaces* **9**, 2200600 (2022).

33. J. van Rijssel, V. F. Peters, J. D. Meeldijk, R. J. Kortschot, R. J. van Dijk-Moes, A. V. Petukhov, B. H. Erne, A. P. Philipse, Size-dependent second virial coefficients of quantum dots from quantitative cryogenic electron microscopy. *Journal of Physical Chemistry B* **118**, 11000–5 (2014).

34. X. Zhang, A. Poniewierski, S. Hou, K. Sozański, A. Wisniewska, S. A. Wieczorek, T. Kalwarczyk, L. Sun, R. Hołyst, Tracking structural transitions of bovine serum albumin in surfactant solutions by fluorescence correlation spectroscopy and fluorescence lifetime analysis. *Soft Matter* **11**, 2512–2518 (2015).

35. A. Testa, M. Dindo, A. A. Rebane, B. Nasouri, R. W. Style, R. Golestanian, E. R. Dufresne, P. Laurino, Sustained enzymatic activity and flow in crowded protein droplets. *Nat Commun* **12**, 6293 (2021).

36. M. Ijavi, R. W. Style, L. Emmanouilidis, A. Kumar, S. M. Meier, A. L. Torzynski, F. H. T. Allain, Y. Barral, M. O. Steinmetz, E. R. Dufresne, Surface tensiometry of phase separated protein and polymer droplets by the sessile drop method. *Soft Matter* **17**, 1655–1662 (2021).

37. S. Hofmann, N. Kedersha, P. Anderson, P. Ivanov, Molecular mechanisms of stress granule assembly and disassembly. *Biochimica et Biophysica Acta (BBA) - Molecular Cell Research* **1868**, 118876 (2021).

38. S. Jain, J. R. Wheeler, R. W. Walters, A. Agrawal, A. Barsic, R. Parker, ATPase-Modulated Stress Granules Contain a Diverse Proteome and Substructure. *Cell* **164**, 487–498 (2016).

39. G. M. Cooper, "Transport of Small Molecules" in *The Cell: A Molecular Approach. 2nd Edition* (Sinauer Associates, 2000; https://www.ncbi.nlm.nih.gov/books/NBK9847/).

40. S. Bröer, A. Bröer, Amino acid homeostasis and signalling in mammalian cells and organisms. *Biochem J* **474**, 1935–1963 (2017).

41. M. A. Olesen, F. Villavicencio-Tejo, R. A. Quintanilla, The use of fibroblasts as a valuable strategy for studying mitochondrial impairment in neurological disorders. *Translational Neurodegeneration* **11**, 36 (2022).

42. A. Ratti, V. Gumina, P. Lenzi, P. Bossolasco, F. Fulceri, C. Volpe, D. Bardelli, F. Pregnolato, A. Maraschi, F. Fornai, V. Silani, C. Colombrita, Chronic stress induces formation of stress granules and pathological TDP-43 aggregates in human ALS fibroblasts and iPSC-motoneurons. *Neurobiology of Disease* **145**, 105051 (2020).

43. T. J. Böddeker, K. A. Rosowski, D. Berchtold, L. Emmanouilidis, Y. Han, F. H. T. Allain, R. W. Style, L. Pelkmans, E. R. Dufresne, Non-specific adhesive forces between filaments and membraneless organelles. *Nat. Phys.* **18**, 571–578 (2022).

44. T. Ohn, N. Kedersha, T. Hickman, S. Tisdale, P. Anderson, A functional RNAi screen links O-GlcNAc modification of ribosomal proteins to stress granule and processing body assembly. *Nat Cell Biol* **10**, 1224–1231 (2008).

45. D. M. Baron, L. J. Kaushansky, C. L. Ward, R. R. K. Sama, R.-J. Chian, K. J. Boggio, A. J. C. Quaresma, J. A. Nickerson, D. A. Bosco, Amyotrophic lateral sclerosis-linked FUS/TLS alters stress granule assembly and dynamics. *Molecular Neurodegeneration* **8**, 30 (2013).

46. Y. Gwon, B. A. Maxwell, R.-M. Kolaitis, P. Zhang, H. J. Kim, J. P. Taylor, Ubiquitination of G3BP1 mediates stress granule disassembly in a context-specific manner. *Science* **372**, eabf6548 (2021).

47. K. A. Ibrahim, K. S. Grußmayer, N. Riguet, L. Feletti, H. A. Lashuel, A. Radenovic, Label-free Identification of Protein Aggregates Using Deep Learning. bioRxiv [Preprint] (2023). https://doi.org/10.1101/2023.04.21.537833.

48. S. Alberti, Small molecules for modulating protein driven liquid-liquid phase separation in treating neurodegenerative disease. 48.

49. M. G. Page, T. Zemb, M. Dubois, H. Cölfen, Osmotic Pressure and Phase Boundary Determination of Multiphase Systems by Analytical Ultracentrifugation. *ChemPhysChem* **9**, 882–890 (2008).

50. X. Xu, G. de With, H. Cölfen, Self-association and gel formation during sedimentation of like-charged colloids. *Materials Horizons* **9**, 1216–1221 (2022).

51. J. Schindelin, I. Arganda-Carreras, E. Frise, V. Kaynig, M. Longair, T. Pietzsch, S. Preibisch, C. Rueden, S. Saalfeld, B. Schmid, J.-Y. Tinevez, D. J. White, V. Hartenstein, K. Eliceiri, P. Tomancak, A. Cardona, Fiji: an open-source platform for biological-image analysis. *Nat Methods* **9**, 676–682 (2012).

52. C. Stringer, T. Wang, M. Michaelos, M. Pachitariu, Cellpose: a generalist algorithm for cellular segmentation. *Nat Methods* **18**, 100–106 (2021).

53. M. Pachitariu, C. Stringer, Cellpose 2.0: how to train your own model. *Nat Methods* **19**, 1634–1641 (2022).